\documentclass[twocolumn,showpacs,preprintnumbers,amsmath,amssymb]{revtex4}

\usepackage{graphicx}
\usepackage{dcolumn}
\usepackage{bm}

\begin{document}

\preprint{APS/123-QED}

\title{Estimation of critical behavior from the density of states in Classical Statistical Models}

\author{A. Malakis}
\altaffiliation[]{Corresponding author: amalakis@cc.uoa.gr}
\author{A. Peratzakis}
\author{N. G. Fytas}
\affiliation{Department of Physics, Section of Solid State
Physics, University of Athens, Panepistimiopolis, GR 15784
Zografos, Athens, Greece}

\date{\today}

\begin{abstract}
We present a simple and efficient approximation scheme which
greatly facilitates extension of Wang-Landau sampling (or similar
techniques) in large systems for the estimation of critical
behavior. The method, presented in an algorithmic approach, is
based on a very simple idea, familiar in statistical mechanics
from the notion of thermodynamic equivalence of ensembles and the
central limit theorem. It is illustrated that, we can predict with
high accuracy the critical part of the energy space and by using
this restricted part we can extend our simulations to larger
systems and improve accuracy of critical parameters. It is
proposed that the extensions of the finite size critical part of
the energy space, determining the specific heat, satisfy a scaling
law involving the thermal critical exponent. The method is applied
successfully for the estimation of the scaling behavior of
specific heat of both square and simple cubic Ising lattices. The
proposed scaling law is verified by estimating the thermal
critical exponent from the finite size behavior of the critical
part of the energy space. The density of states (DOS) of the
zero-field Ising model on these lattices is obtained via a
multi-range Wang-Landau sampling.
\end{abstract}

  \maketitle

\section{Introduction}
In the past half-century, importance sampling in the canonical
ensemble and especially the Metropolis method and its variants was
the main tool in condensed matter physics, mainly for the study of
critical phenomena
\cite{metro53,glaub63,bortz75,binde77,newma99,landa00}. However,
this standard approach has two serious disadvantages. The
partition function of the statistical model is not an output of
such calculations and in many cases importance sampling is trapped
for significant time in valleys of rough free energy landscape.
Over the last decade, there have been a number of interesting
approaches addressing to these problems
\cite{newma99,landa00,lee93,berg92,ferre98,lima00,wang02,wang99,wang01,wang01b}.
Recently efficient methods that directly calculate the density of
states (DOS), or the spectral degeneracy, of classical statistical
models have been developed. A few remarkable examples are the
entropic \cite{newma99,lee93}, multicanonical \cite{berg92},
histogram and broad histogram \cite{ferre98,lima00}, transition
matrix \cite{wang02,wang99} and Wang-Landau
\cite{wang01,wang01b,schul03,schul01,landa02} methods. The above
methods are thought to be the most promising ones for the
application of finite-size analysis in data with higher accuracy.
It is well known \cite{landa00} that finite-size analysis is very
sensitive to simulational errors and in most cases the asymptotic
analysis may become a notorious task, due to the fact that these
errors may ``interfere'' with unknown correction terms
\cite{landa00}.

In this paper we concentrate on the Wang-Landau method
\cite{wang01,wang01b,schul03,schul01,landa02}. Noteworthy that
this method was applied by these authors on the two-dimensional
Ising model producing the density of states even on lattices, as
large as, $256\times256$. This was achieved by a multi-range
algorithm in which independent random walks were used for
different energy subintervals and the resultant pieces were then
combined to obtain the density of states. The possibility of
producing accurate estimates for critical-point anomalies on large
lattices establishes, at least it is hoped, Wang-Landau method,
and similar techniques, as new and important tools for evaluating
equilibrium properties of models showing complex properties of
substances, such as systems with competing interactions and spin
glass models. Therefore, it is of interest to understand how we
can implement these methods in the best way for the extraction of
critical parameters using the finite-size scaling theory.  This
paper considers an important aspect of this problem and aims to
introduce a simple and practical route, through which we can
substantially improve both accuracy and efficiency of the above
methods. It will be shown that only a relatively small part of
spectral degeneracies is needed in order to obtain a good
estimation of critical properties. This part of the total energy
range can be easily identified. In Section II we present an
outline of the method which one can use to identify the subspace
of the energy space that determines the specific-heat peak
behavior. Also a very brief description of the multi-range
Wang-Landau method is presented. In Section III the proposed
method is tested for the plane square Ising lattice where the
finite-size scaling behavior is known from the work of A.
Ferdinand and M. Fisher \cite{ferdi69}. In Section IV we discuss
the critical-point specific heat anomaly behavior of the simple
cubic Ising lattice. We present estimates for the critical
temperature and the associated critical exponents and compare our
results with existing estimates. It is shown that the extension of
the used critical energy subspace scales as predicted in the
theory presented in Section II and this provides an independent
estimation of the ratio $\alpha/\nu$  of critical exponents. We
summarize our results and conclusions in Section V.

\section{Estimation of the critical part of the energy space}
Let us consider the zero-field Ising model on the $L\times L=N$
square and the $L\times L\times L=N$ cubic lattices:
\begin{equation}
\label{eq:1}
H=-J\sum_{<ij>}S_{i}S_{j},\;\;\;\;S_{i}=\pm1,\;\;\;\;i=1,2,...,N
\end{equation}
The behavior of finite systems near the infinite lattice critical
temperature $T_{c}$ can be described by finite-size scaling
theory\cite{fishe71,privm90,binde92}. For the three dimensional
Ising model the maxima of the finite-size specific heats
$C^{\ast}_{L}$  are expected to scale as:
\begin{subequations}
\label{eq:2}
\begin{equation}
\label{eq:2a} C^{\ast}_{L}=c+bL^{\alpha/\nu}(1+...)
\end{equation}
Where $\alpha$ and $\nu$ are the critical exponents of the
specific heat and the correlation length, respectively. For the
square Ising lattice the (logarithmic) scaling of the maxima of
the finite-size specific heats $C^{\ast}_{L}$  is known from the
work of A. Ferdinand and M. Fisher \cite{ferdi69} and will be
considered in the next section. The shift of the
``pseudo-critical'' temperatures $T^{\ast}_{L}$ (defined by the
location of the specific-heat peaks) is described by a similar
power law for both square and cubic Ising lattices:
\begin{equation}
\label{eq:2b} T^{\ast}_{L}=T_{c}+cL^{-1/\nu}(1+...)
\end{equation}
\end{subequations} Given an approximation for the density of
states, $G(E)$, obtained for instance via the  Wang-Landau method,
the specific heat at any temperature can be estimated and thus
the``pseudo-critical'' temperature $T_{L}^{\ast}$  and the maximum
of the specific heat are easily obtained. Therefore, applying such
a method to finite systems, we can accumulate data and through the
finite size scaling mechanism extract the asymptotic critical
behavior. Of course, this has been done in the past using the more
traditional Monte Carlo methods (particularly importance sampling
techniques). Using these later methods we have to simulate the
system in a range of temperatures around the ``pseudo-critical''
temperature $T^{\ast}_{L}$. For each such temperature we have to
perform our simulations for a suitably long period of time for
``equilibration'' and then make a large number of independent
measurements for averaging. In effect, this usually means a very
large number (several millions) of Monte Carlo steps determined
mainly by the critical slowing down phenomenon
\cite{newma99,landa00}. On the other side, using the Wang-Landau
method one has ``at once'' an approximation of the specific heat
at any temperature and thus, as mentioned above, the
``pseudo-critical'' temperature $T^{\ast}_{L}$ and the maximum of
the specific heat are easily obtained. However, executing a
Wang-Landau random walk process in the total energy space can also
be time-consuming and moreover the almost unavoidable multi-range
algorithm will definitely introduce some ``uncontrollable''
errors. These errors are ``histogram errors'' which may propagate
and amplify through the process of connecting the energy ranges in
a multi-range approach. Note that by applying separately a
histogram flatness criterion (such as (\ref{eq:11})) in each
energy range does not produce necessarily the same level of
flatness in the total energy range. It is therefore profitable if
we can estimate, with the same or even better accuracy, the
specific heat peaks using only a small part of the energy space.

In order to proceed we express the value of the specific heat, at
any temperature, with the help of the usual statistical sum
$(k_{B}=1)$:
\begin{subequations}
\label{eq:3}
\begin{widetext}
\begin{equation}
\label{eq:3a} C_{L}(T)=N^{-1}T^{-2}\left\{
Z^{-1}\sum^{E_{max}}_{E_{min}}E^{2}\exp[{\Phi(E)}]-\left(Z^{-1}\sum^{E_{max}}_{E_{min}}E\exp[{\Phi(E)}]\right)^{2}\right\}
\end{equation}
\end{widetext}
where $\Phi(E)$ and the microcanonical entropy $S(E)$ are defined
by:
\begin{equation}
\label{eq:3b}
\Phi(E)=S(E)-\beta E,\;\;\;S(E)=\ln{G(E)}
\end{equation}
while the function Z by:
\begin{equation}
\label{eq:3c}
Z=\sum_{E_{min}}^{E_{max}}\exp{[\Phi(E)]}
\end{equation}
\end{subequations} Note that $Z$ is the partition function in case
one uses the total energy spectrum and $G(E)$ is properly
normalized. The above expressions give, in fact, an approximation
for the values and the maximum of the specific heat since
Wang-Landau simulations provide us with an approximate DOS $G(E)$.

Now let $\widetilde{E}$ denote the value of energy producing the
maximum term in the sum (\ref{eq:3c}) of the (partition) function
at a temperature of our interest. Eventually, we will concentrate
on the ``pseudo-critical'' temperature $T_{L}^{\ast}$ or some
temperatures close to this (for instance the exact critical
temperature $T_{c}$, whenever this temperature is known). We may
define a set of approximations to the specific-heat values by
restricting the statistical sums in (\ref{eq:3}) to energy ranges
around this value. We define the following energy sub-ranges of
the total energy range ($E_{min},E_{max}$):

\begin{equation}
\label{eq:4} (\widetilde{E}_{-},\widetilde{E}_{+}),\;\;\;\;
\widetilde{E}_{\pm}=\widetilde{E}\pm \Delta^{\pm},\;\;\;\;\;
\Delta^{\pm}\geq0
\end{equation}
Accordingly, the value of the specific heat at the temperature of
interest (for instance, the value of the peaks at
the``pseudo-critical'' temperature $T^{\ast}_{L}$ ) are
approximated by:
\begin{subequations}
\label{eq:5}
\begin{widetext}
\begin{equation}
\label{eq:5a}
C_{L}(\widetilde{E}_{-},\widetilde{E}_{+})\equiv
C_{L}(\Delta^{\pm})=N^{-1}T^{-2}\left\{\widetilde{Z}^{-1}\sum_{\widetilde{E}_{-}}^{\widetilde{E}_{+}}E^{2}\exp{[\widetilde{\Phi}(E)]}\right.\\
\left.-
\left(\widetilde{Z}^{-1}\sum_{\widetilde{E}_{-}}^{\widetilde{E}_{+}}E\exp{[\widetilde{\Phi}(E)]}\right)^{2}\right\}
\end{equation}
\end{widetext}
where
\begin{equation}
\label{eq:5b} \widetilde{\Phi}(E)=[S(E)-\beta
E]-\left[S(\widetilde{E})-\beta \widetilde{E}\right]
\end{equation}
and
\begin{equation}
\label{eq:5c}
\widetilde{Z}=\sum_{\widetilde{E}_{-}}^{\widetilde{E}_{+}}\exp{[\widetilde{\Phi}(E)]}
\end{equation}
\end{subequations}
Depending on the extension of the sub-ranges used in (\ref{eq:5})
the above sequence may give good approximations to the specific
heat values.

As mentioned earlier, finite-size analysis depends on the accuracy
of the finite lattice data, in a very sensitive way. Therefore,
one may question the utility of introducing further approximations
to an already approximate scheme. To answer this we demand that
the new errors are much smaller than the already existing ones
from the DOS approximation. Since by definition
$\widetilde{\Phi}(E)$ is negative we can easily see that for large
lattices ``extreme'' values of energy (far from $\widetilde{E}$)
will have an extremely small contribution to the statistical sums
since these terms decay exponentially fast with respect to the
distance from $\widetilde{E}$. It follows that, if we request a
specified accuracy (assume that our approximations satisfy some
strict criteria), then we may greatly restrict the necessary
energy range, in which DOS should be sampled. If this is so, then
we not only reduce computer time for the calculation of the
approximate DOS, but we also improve accuracy. Indeed, by
restricting the energy space we should expect a minimization of
``Wang-Landau errors'' even in cases where a multi-range approach
is used.

To make this idea concrete we demand that the relative errors
introduced by the restriction (\ref{eq:5}) are smaller than a
given number $\textit{r}$. Moreover we assume that these relative
errors are considerably smaller than those produced by the
Wang-Landau scheme on the values of the specific heat. This
restriction is well defined if we know the exact DOS for a finite
system. Given any small number and the exact DOS one can easily
calculate the minimum energy subspace (MES), compatible with the
above restriction. An algorithmic approach is described below in
equations (\ref{eq:9}). The resulting subspace (its end-points and
its extension) depends, of course, on the temperature, on the
value of the small parameter $\textit{r}$, and on the lattice
size. We write for its extension, $\Delta\widetilde{E}$:
\begin{eqnarray}
\label{eq:6}
&\Delta\widetilde{E}\equiv\Delta\widetilde{E}(T,\textit{r},L)\equiv
min(\widetilde{E}_{+}-\widetilde{E}_{-}):\nonumber\\
&\mid\frac{C_{L}(\Delta^{\pm})}{C_{L}}-1\mid\leq\textit{r}
\end{eqnarray}

Closely related to the notion of the thermodynamic equivalence of
ensembles and to the central limit theorem is here the idea that,
for any temperature, the extension of the energy subspace
determining the behavior of the system is much smaller than the
total energy range:
$\Delta\widetilde{E}(T,\textit{r},L)\ll(E_{max}-E_{min})$. Thus,
our proposition of using Wang-Landau sampling in the critical MES
is quite obvious. We should expect that the extension of the
above-defined restricted part of the energy space would be of the
same order with the standard deviation of the energy distribution
at any temperature. Therefore, we assume that given a small
constant value for $\textit{r}$:
\begin{equation}
\label{eq:7} \Delta\widetilde{E}\propto\sigma_{E}=\sqrt{NT^{2}C}
\end{equation}
From the central limit theorem we know that, far from the critical
point, the energy distribution approaches a Gaussian distribution
and the energy subspace determining all thermodynamic properties
is mostly of the order of $\sqrt{N}$. Close to a critical point
the order of the extension of MES is not known, but assuming
thermodynamic equivalence of ensembles one should expect the
extension of critical MES to be $\ll N$ . Although the energy
distribution will diverge from the Gaussian, it still seems
reasonable to describe the extensions of the critical MES,
$\Delta\widetilde{E}^{\ast}=\Delta\widetilde{E}(T^{\ast}_{L},\textit{r},L)$
by (\ref{eq:7}). Therefore, using (\ref{eq:2a}) we may conclude
that these extensions, as well as the values
$\Delta\widetilde{E}_{c}=\Delta\widetilde{E}(T_{c},\textit{r},L)$,
should scale as:
\begin{equation}
\label{eq:8} \frac{\Delta\widetilde{E}^{\ast}}{L^{d/2}}\approx
L^{\frac{\alpha}{2\nu}}
\end{equation}

In order to obtain the MES from the exact DOS or from a given
approximation of DOS, $G(E)$, we define  successive `minimal'
approximations to the specific-heat values:
\begin{subequations}
\label{eq:9}
\begin{eqnarray}
\label{eq:9a}
&C_{L}(j)\equiv C_{L}(\Delta_{j}^{-},\Delta_{j}^{+}),\nonumber\\
&\Delta^{\pm}_{j+1}=\Delta^{\pm}_{j}\pm\theta^{\pm}_{j+1},\;\;\
\Delta_{1}^{\pm}=0,\;\;\;\ j=1,2,...
\end{eqnarray}

One of the above $\theta$-increments is chosen to be 1 and the
other 0 according to which side of $\widetilde{E}$ is producing at
the current stage the best approximation:
\begin{eqnarray}
\label{eq:9b}
&(\theta^{+}_{j+1}=1,\theta^{-}_{j+1}=0)\;\Leftrightarrow\nonumber\\
&\mid
C_{L}-C_{L}(\Delta_{j}^{-},\Delta^{+}_{j}+1)\mid\leq\nonumber\\
&\leq\mid C_{L}-C_{L}(\Delta_{j}^{-}+1,\Delta_{j}^{+})\mid\;
\end{eqnarray}
\begin{eqnarray}
\label{eq:9c}
&(\theta^{+}_{j+1}=0,\theta^{-}_{j+1}=1)\;\Leftrightarrow\nonumber\\
&\mid C_{L}-C_{L}(\Delta_{j}^{-},\Delta^{+}_{j}+1)\mid>\nonumber\\
&>\mid C_{L}-C_{L}(\Delta_{j}^{-}+1,\Delta_{j}^{+})\mid
\end{eqnarray}
Accordingly the sequence of relative errors for the specific-heat
values ($\textit{r}_{j}$) is given by:
\begin{equation}
\label{eq:9d}
\textit{r}_{j}=\mid\frac{C_{L}(j)}{C_{L}}-1\mid
\end{equation}
\end{subequations}
We now fix our requirement of accuracy by specifying a particular
level of accuracy for all finite lattices. In effect, we define
the (critical) MES as the subspace centered at $\widetilde{E}$
($\widetilde{E}^{\ast}$) corresponding to the first member of the
above sequence (\ref{eq:9}) satisfying:
$\textit{r}_{j}\leq\textit{r}$. Demanding the same level of
accuracy for all lattice sizes, we produce a size-dependence on
all parameters of the above energy ranges. That is we should
expect that the ``center'' $\widetilde{E}(T,L)$ and the end-points
$E_{-}(T,r,L)$, $E_{+}(T,r,L)$ of the (critical) MES are all
functions of L. In particular the extensions
$\Delta\widetilde{E}^{\ast}=\Delta\widetilde{E}(T_{L}^{\ast},\textit{r},L)$
of the critical MES should obey the scaling law (\ref{eq:8}). It
is therefore possible to find approximations of these functions
using the total energy range for small lattices and then
extrapolate to estimate the critical MES for larger lattices. A
slightly wider energy subspace is easily predicted by
extrapolating from smaller lattices. Furthermore, working in a
wider range we can use our approximate DOS to have a very good
approximation of the critical MES and thus check the validity of
the proposed scaling law (\ref{eq:8}). This is possible, because,
the approximations are expected to obey an exponentially fast
convergence outside the energy range centered in $\widetilde{E}$.
We may also apply the above r-depended scheme for several values
of the accuracy parameter $\textit{r}$. Once the accuracy criteria
have been satisfied for a given value of $\textit{r}$ and the
energy range is wide enough to accurately estimate the
corresponding CrMES, we can also estimate the extensions of CrMES
for any larger value of the parameter $\textit{r}$.

\begin{figure*}
\includegraphics*[width=12 cm]{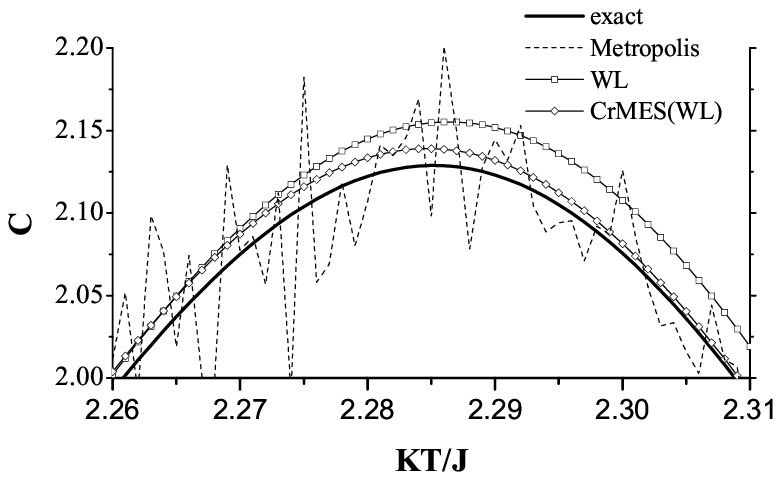}
\caption{\label{fig:fig1} Comparative diagram showing the exact
specific heat (in units of $k_{B}$) for a $50\times50$ square
Ising lattice (solid line) and approximate curves corresponding to
(i) Metropolis Importance sampling. Average behavior over $20$
samples. For each temperature we have used after equilibration
about $\sim 10^{4}$ Monte Carlo sweeps for averaging. (ii)
Wang-Landau multi-range sampling ($4$ independent random walks
transversing the total energy range. (iii) Wang-Landau multi-range
sampling applied to the critical energy subspace (CrMES) (i.e.
$20$ independent random walks ran in a slightly wider energy range
of extension $\sim 450$ energy levels). Note that (ii) and (iii)
have approximately the same time requirements. Obviously the
improved accuracy of the proposed scheme is apparent mainly
because, for the same available time, one can perform more
independent WL random walks.}
\end{figure*}

Let us now briefly discuss the main points of our implementation
of the Wang-Landau method. For the application of the algorithm in
multi-range approach we follow the description of  Schulz et al.
$2003$, i.e., whenever the energy-range is restricted we use the
updating scheme $2$ described in that paper. Consider the
restriction of the random walk in a particular energy-range
$I=[E_{1},E_{2}]$ and assume that the random walk is at the border
of the range I. Then, the next spin-flip attempt is determined by
the modified Metropolis acceptance ratio:
\begin{equation}
\label{eq:10} A=\left\{\begin{array}{rr}
\min\{1,G(E)/G(E+\Delta E)\}, & \mbox{$(E+\Delta E)\in I$} \\
0, & \mbox{$(E+\Delta E)\not\in I$} \end{array}\right.
\end{equation}
The random walk is not allowed to move outside of the energy
range, and we always update the histogram value $H(E)\rightarrow
H(E)+1$ and the DOS value $G(E)\rightarrow
G(E)^{\ast}\textit{f}_{j}$ after a spin-flip trial. Here, of
course, $\textit{f}_{j}$ is the value of the Wang-Landau
modification factor \cite{wang01,wang01b,schul03,schul01,landa02},
at the $j^{th}$ iteration, in the process
($\textit{f}\rightarrow\textit{f}^{\, 1/2}$) of reducing its value
to 1, where the detailed balance condition is satisfied. In all
our simulations the Wang-Landau modification (or the control
parameter), was chosen to have the initial value:
$\textit{f}_{j=1}=e\approx2.71828...$. When starting a new
iteration the control parameter is changed according to
$\textit{f}_{j+1}=\sqrt{\textit{f}_{j}}, j=1,2,...$ \cite{wang01}.
Also, we use the following criterion for the histogram flatness:
\begin{equation}
\label{eq:11} \frac{maxH(E)-minH(E)}{maxH(E)}\leq0.05
\end{equation}
Using a multi-range approach, we divide the total energy range or
the expected MES in several subintervals overlapping in one or
several points at their ends. These subintervals can be then
joined at the end to obtain the DOS in the range of interest. In
joining two neighboring subintervals the degeneracies in one of
the two have to be adjusted so that its endpoint degeneracies
conform to the corresponding degeneracies of the neighboring
interval. Obviously, this is a process that may propagate
``histogram errors'', but from the description above it is
apparent that one can arrange this process to leave unchanged the
``central subinterval degeneracies''. Since this subinterval can
be chosen to have its center close to the energy value
$\widetilde{E}(T^{\ast}_{L})$, this choice will be optimal and it
will produce relatively small errors. Usually, when one is
sampling the total DOS, a normalization condition is applied. This
condition may concern the ground state degeneracy, or the total
number of states of the system, or even some convenient
combination of known degeneracies. However, normalization does not
effect the values of the specific heat, so we may conveniently
choose $G(\widetilde{E}')=1$, where $\widetilde{E}'$ is an initial
guess for $\widetilde{E}(T_{L}^{\ast})$ which serves as the center
of the ``central subinterval''. Furthermore, since the central
subinterval is the most influential in the determination of the
specific-heat peak, we have chosen the subintervals to have
varying lengths(of the order of $50-160$ energy levels) with the
largest to be the central subinterval ($100-160$ energy levels).
Finally, in each case we observed the behavior in a sample of
several independent runs and the j-iteration process
($\textit{f}_{j+1}=\sqrt{\textit{f}_{j}}, j=1,2,...$) is carried
out, until fluctuations around a `mean' for the specific-heat peak
are obtained. In almost all cases, this occurred in the range
between $20-26$ Wang-Landau iterations for the modification
factor. Fig.\ref{fig:fig1} shows the application of Monte Carlo
approaches to the specific heat peak for a $50\times50$ square
Ising lattice. The traditional Metropolis importance sampling, the
Wang-Landau multi-range sampling of the total DOS and the proposed
in this paper ``critical minimum energy subspace (CrMES)''
sampling are compared to the exact specific-heat peak.

\section{A Test Case: The Square Ising Lattice}
A. Ferdinand and M. Fisher \cite{ferdi69} have asymptotically
analyzed the critical point anomaly of a
$\textit{m}\times\textit{n}$ plane square Ising lattice with
periodic boundary conditions. In that paper, which has been one of
the most influential papers in the development of finite-size
scaling theory, an explicit expansion for the specific heat close
to the critical point is described. We shall use their asymptotic
expansion to test our simulational data obtained via Wang-Landau
scheme in the CrMES. This is a first examination of our proposal
for estimating the critical behavior through a ``CrMES Wang-Landau
scheme''. Furthermore, by performing the Wang-Landau random walk
in a slightly wider range than the CrMES, we can estimate the
finite-size extensions of the CrMES and explore the possibility of
estimating the critical behavior using the scaling law proposed in
(\ref{eq:8}).

Since for the two-dimensional case the exact critical temperature
$T_{c}$ is known, it is useful to apply the ``CrMES Wang-Landau
scheme'' for both the exact critical temperature and the
``pseudo-critical'' temperature $T^{\ast}_{L}$. Actually, this
means that we have to run the Wang-Landau random walk in a wider
range. As an example, let us give details for the case of a
$50\times50$ lattice. Counting the energy levels as
$ie(E)=(E+2N)/4+1$, that is starting the enumeration from the
ground state, the energy levels corresponding to the ``centers''
for the two temperatures of interest are
$ie(\widetilde{E}(T_{c}))=354$ and
$ie(\widetilde{E}(T_{L}^{\ast}))=378$. The corresponding
extensions for a chosen level of accuracy $\textit{r}=10^{-6}$ are
$392$ and $394$ respectively. Note that these `extensions' are
measured in terms of the convenient counting integer variable
$ie(E)$. The extensions of the critical ranges are of the same
order but their ``centers'' do not coincide (reflecting the shift
of the ``pseudo-critical'' temperature). Their displacement is
$24$ energy levels, so in order to achieve (for the specific-heat
approximation (\ref{eq:5})) a relative error of the order of
$\textit{r}=10^{-6}$, we should execute the Wang-Landau random
walk in a range of the order of $420$ $(394+24)$ energy levels. If
we furthermore request to accurately estimate the extensions of
the corresponding MES, we should consider a slightly wider (from
each side) range, which in the case of a $50\times50$ lattice
modifies the required range of the order of $450$ energy levels.
Thus, the number of energy levels for the Wang-Landau random walk
is greatly reduced, since this number should be compared to a
total of $2500$ energy levels for the $50\times50$ lattice. A
practical method for guessing the wider range necessary for an
accurate estimation of the extensions of the CrMES may be as
follows. We can easily devise a ``self-consistent'' test to
inspect from the derived DOS (in the wider range) whether or not
the estimated critical extensions are completely determined in
this wider range. This is easily accomplished by using
successively increasing ranges, starting from the estimated CrMES
ranges, and observing the variation in the estimated extensions as
the range grows up to the final wider version. Because of the
exponential convergence mentioned earlier, this procedure will
converge very fast. Thus, one can manage to know after his runs
whether the originally ``guessed'' (or estimated through an
extrapolation scheme) range was large enough to produce accurately
the extensions of CrMES. After a successful run for a given
lattice size we may know, if we wish, to what percentage was
necessary to increase the CrMES (from each side) in order to
accurately estimate its extension. This information may be then
used for extrapolation to larger lattice sizes.

Table \ref{tab:table1} presents the ``pseudo-critical''
temperatures, the corresponding values of the specific heats as
well as the values of the specific heats at the exact critical
temperature for lattice sizes $L=10-100$. The results for
$L=10-50$ were obtained using the exact DOS derived by executing
the algorithm provided by \cite{beale96}, while the estimates for
the larger lattices by the proposed CrMES Wang-Landau scheme. For
the lattice $50\times50$ both exact and approximate results are
shown for comparison. For each lattice size we have considered
$20$ random walks in order to improve statistics. To obtain
estimates of the errors, the specific-heat values were calculated
separately from the DOS of each random walk (thus taking
afterwards suitable averages), but also from the averaged DOS of
the sample. We have also observed the variation of the estimated
parameters as a function of the order of Wang-Landau iteration in
the process of reducing the modification factor. Although we do
not know any general criterion for an `optimum' estimation using
the Wang-Landau technique, we think it is a good practice to
observe the variation of the estimated parameters as we proceed in
higher orders of the approximate scheme. The errors given in
brackets reflect the order of the standard deviation of averaging
the separate walk estimates. Of course, using groups of random
walks one may reduce these errors but this will not effect the
estimated mean. The estimates given in tables are averages of the
two processes, i.e. mean values of the estimates obtained from the
separate walks and the estimates obtained from the averaged (over
a sample of $20$ random walks) DOS. The values of the $24^{th}$
Wang-Landau iteration were used in most cases, but the behavior
was observed for the $20-26^{th}$ iterations.

Let us now see how one could try from these data to estimate the
critical parameters assuming that, at least, the leading behavior
is known. From the work of A. Ferdinand and M. Fisher
\cite{ferdi69} we know that close to the critical point:
\begin{equation}
\label{eq:12}
C_{L}(T)=A_{0}\ln{L}+B(T)+B_{1}(T)\frac{\ln{L}}{L}+B_{2}(T)\frac{1}{L}+...
\end{equation}
where the critical amplitude $A_{0}$ and the first $B$
coefficients are given in \cite{ferdi69} for both the exact
critical and ``pseudo-critical'' temperatures (see also bellow).
We try to fit the data of Table \ref{tab:table1} for
$C_{L}(T_{c})$ and $C_{L}(T_{L}^{\ast})$ to the above expansion
and reproduce the correct amplitudes. However, it is well known
\cite{landa00,ferre91} that including many independent correction
terms, even when high-quality data are available is not a
suggested procedure, unless we have almost exact data up to very
large lattices. In all other cases, it seems that the best one can
do is to start with (or search for) the dominant correction term.
Therefore, in the present case we consider only the first two
terms in the above expansion (setting the other terms zero) and
pay attention in estimating the critical amplitude $A_{0}$(mainly)
and the constant B-contribution.

Fitting the finite size values (exact and approximate, where for
the sizes $L=54-100$ the Wang-Landau data of Table
\ref{tab:table1} are used) of the specific heat at the
corresponding ``pseudo-critical'' temperatures for sizes $10-50,
L=10-100$ and $L=50-100$ we obtain the following estimates for the
critical amplitude and the constant B-contribution:
\begin{eqnarray}
\label{eq:13}
&L&=10-50:\ \;\ A_{0}\cong0.509(1),\;\ B^{\ast}\cong0.140(4)\nonumber\\
&L&=10-100:\;\ A_{0}\cong0.504(1),\;\ B^{\ast}\cong0.154(5)\nonumber\\
&L&=50-100:\;\ A_{0}\cong0.494(5),\;\ B^{\ast}\cong0.198(2)\nonumber\\
\end{eqnarray}
Similarly, applying the same fittings for the specific-heat data
at the exact critical temperature we find:
\begin{eqnarray}
\label{eq:14}
&L&=10-50:\ \;\ A_{0}\cong0.503(1),\;\ B_{c}\cong0.104(2)\nonumber\\
&L&=10-100:\;\ A_{0}\cong0.499(2),\;\ B_{c}\cong0.116(6)\nonumber\\
&L&=50-100:\;\ A_{0}\cong0.494(8),\;\ B_{c}\cong0.138(33)\nonumber\\
\end{eqnarray}

These estimates are to be compared with the values given in A.
Ferdinand and M. Fisher \cite{ferdi69}:
\begin{eqnarray}
\label{eq:15}
&A_{0}=0.494358...,\;\;\ B^{\ast}=B(T^{\ast}_{L})=0.201359...,\nonumber \\
&B_{c}=B(T^{\ast}_{c})=0.138149...
\end{eqnarray}
As expected the inclusion of data for larger sizes improves the
estimates and one can see that the improvements are in the right
direction. Hence, one can further refine the estimates by using
more sophisticated extrapolation schemes and possibly by taking
into account data for even larger lattices.

Let us now turn our attention to the verification of the proposed
in (\ref{eq:8}) scaling law for the extensions of the CrMES. Table
II presents the extensions of the MES for both the exact critical
temperature and the ``pseudo-critical'' temperature for all sizes
considered in Table \ref{tab:table1}. Again, the extensions
presented for sizes $L=10-50$ were obtained using the exact DOS
while the estimates for the larger lattices were obtained by the
proposed CrMES Wang-Landau scheme and for the $50\times50$ case
both exact and approximate results are shown as a comparison. A
striking observation concerns the errors of these artificially
constructed parameters. In fact for very large lattice sizes there
are relatively small errors, while for moderate sizes there are no
errors at all. Indeed, despite the fact that the reported errors
were obtained in the same way as in the case of the specific heat
values, the relative errors of the extensions are smaller by a
factor of $10$ for the largest lattice size used $L=100$. The
``center'' of the CrMES fluctuates from walk to walk due to the
approximate DOS produced by the Wang-Landau scheme. However, the
errors in determining these ``central'' points are in general
greater than the errors in determining the extensions of MES.
Table \ref{tab:table2} contains the extensions of the CrMES for
three different levels of accuracy specified by
$\textit{r}=10^{-3}, 10^{-4}$ and $\textit{r}=10^{-6}$. At this
point we note that even the largest value of $\textit{r}$
determining the accuracy level in (\ref{eq:6}) is smaller than the
relative errors produced by the Wang-Landau technique. The
proposed in (\ref{eq:5}) approximation, by restricting the energy
space will not introduce errors outside the limits of the
Wang-Landau accuracy. As pointed out our calculations were done in
sufficiently wide ranges so that the extensions of the CrMES were
accurately estimated. Note that if our runs were performed in a
wide enough range, sufficient to accurately estimate the
extensions of MES for say the third criterion, then this range
would be sufficiently wide for the estimation of the extensions
corresponding to any larger value of $\textit{r}$.

\begin{figure*}
\includegraphics*[width=12 cm]{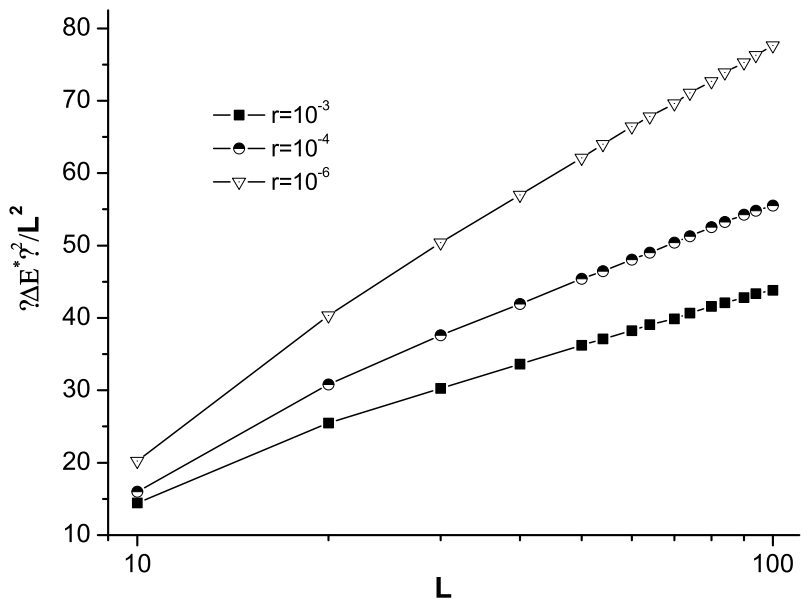}
\caption{\label{fig:fig2}Demonstration of the logarithmic scaling
law (\ref{eq:16}) of the CrMES extensions for the square Ising
model, shown for the three levels of accuracy chosen. Note that
the extensions $\Delta E^{\ast}$ are defined to be dimensionless
as discussed in Section III.}
\end{figure*}

Trying to fit these extensions to an asymptotic expansion of the
form (\ref{eq:12}) we find that the dominant correction is the
third term. Thus we use the following formula:
\begin{equation}
\label{eq:16}
\Psi(\textit{r})\equiv\left(\frac{\Delta\widetilde{E}(\textit{r})}{L^{d/2}}\right)^{2}\approx
A(\textit{r})\ln{L}+B_{1}(\textit{r})\frac{\ln{L}}{L}
\end{equation}
Table \ref{tab:table3} gives the estimates for the above
amplitudes for sizes $L=10-50$, $L=10-100$ and $L=50-100$. Also
Fig.\ref{fig:fig2} shows the behavior of theses extensions versus
lattice size. As was expected on physical grounds, the extensions
of CrMES follow the same asymptotic law with the specific heat in
the critical region and provide a new independent method of
estimating critical behavior via the finite size scaling analysis.
Since, $\Delta\widetilde{E}_{c}\cong\Delta\widetilde{E}^{\ast}-2$
for all lattice sizes, the extensions of the CrMES at the exact
critical temperature follow the same scaling law. We end this
section by noting that one can use the data in Table
\ref{tab:table1} to estimate from the law (\ref{eq:2b}) the
critical exponent $\nu$ and the critical temperature $T_{c}$.

\section{The Three - Dimensional Ising Model}
Despite the intense effort made over the last decades, the
three-dimensional Ising model has defied exact solution
\cite{domb60,fishe67} and though it has been investigated
extensively by various numerical methods, is still a matter of
sophisticated numerical analysis
\cite{nicke80,brezi74,guill80,guill87,pawle84,blote89,blote89b,liu89,landa76,barbe85,paris85,hoogl85,bhano86,chen82,guttm94,koles94,nicke90,nicke91,adler83,talap96,blote95,blote96b,deng03,blote97,hanse91,buter97,guida98,garci,buter02}.
The critical properties of the model, i.e., the critical
temperature $T_{c}$, the thermal and magnetic scaling exponents
$y_{t}$ and $y_{h}$, and also the leading thermal irrelevant
exponent $y_{i}$ seem to be known with good accuracy
\cite{deng03}. However, the absence of exact results creates, at
least in principle, a motive for disagreements
\cite{garci,deng03}. For many years reliable estimates for $T_{c}$
and the critical exponents have been obtained by series-expansion
data, $\varepsilon$-expansion studies, Monte Carlo renormalization
group studies and the coherent anomaly method
\cite{nicke80,brezi74,guill80,guill87,pawle84,blote89,blote89b,liu89,chen82,guttm94,koles94,nicke90,nicke91,adler83,buter02}.

The traditional Monte Carlo sampling, importance sampling and
histogram techniques, have been used also to investigate the
three-dimensional Ising model
\cite{ferre91,landa76,barbe85,paris85,hoogl85,bhano86,blote95,blote96b,deng03}
but only recently \cite{deng03} such studies have provided
accurate estimates of the critical exponents. There are two
reasons for the modest accuracy obtained in these Monte Carlo
simulations. Firstly, extended runs are necessary to reduce the
systematic and statistical errors, which arise due to the finite
number of samples taken. Secondly, corrections to scaling are much
more important in three than in two dimensions. The leading
irrelevant thermal exponent for the three-dimensional Ising model
has the value $y_{i}=-0.821(5)$ \cite{deng03} and this means that
corrections decay relatively slowly. The two effects of finite
sampling time and finite system size become intertwined and
jeopardize the finite size scaling analysis. In particular, it has
been very difficult to accurately estimate the thermal critical
exponent from finite size scaling analysis of Monte Carlo
specific-heat data  close to the pseudo-critical temperatures.

Blote et al \cite{blote95} have presented an extensive Monte Carlo
simultaneous analysis of three cubic Ising models belonging to the
same universality class. Their data were obtained by several
``cluster'' algorithms and their analysis included a finite-size
scaling study for the specific heat anomaly of the simple cubic
Ising model (section $5.2$ in \cite{blote95}). Furthermore, in a
recent analogous study Deng and Blote \cite{deng03} proposed a
different ``better'' route for the estimation of the thermal
exponent. In this latter study, a quantity ($Q_{p}$) that
correlates the magnetization distribution with the energy density
\cite{deng03}, which has a stronger divergence with respect to the
system-size, is used. In general, it appears that the traditional
route for the estimation of the thermal critical exponent, via
specific-heat data, has been overlooked over the years because of
the problems faced in trying to fit these data. This is entirely
understandable by comparing the high accuracy obtained in the
recent paper by Y. Deng and H. Blote \cite{deng03}
($y_{t}=1.5868(3)$), with the modest estimate ($y_{t}=1.60(2)$) in
Blote et al \cite{blote95}. In view of this situation, it is of
interest to apply our proposal for estimation of the DOS via a
Wang-Landau random walk in the CrMES and study again the so
produced numerical data for the specific-heat peaks. Furthermore,
it is most appealing to examine whether the data for the
$\textit{r}$-dependent extensions of the CrMES, give when
subjected to finite-size analysis, estimates in agreement with the
already known values of the thermal critical exponent.

We may, following Blote et al \cite{blote95}, use an expansion for
the specific-heat values close to the critical point of the form:
\begin{widetext}
\begin{equation}
\label{eq:17} C_{L}=L^{2y_{t}-d}\left(
q_{o}+q_{1}(K-K_{c})L^{y_{t}}\right)+p_{o}+rL^{2y_{t}-d+y_{i}}+s_{o}L^{y_{t}-d}
\end{equation}
\end{widetext}
In this expansion the renormalization group behavior of the free
energy with a scale factor ($l$) has been used. Moreover, the
existence of an irrelevant field has been assumed and some terms
from the more general expansion have been omitted as dominated by
the correction terms included in (\ref{eq:17}) (see equation
$(21)$ and discussion in \cite{blote95}). Blote et al
\cite{blote95} used a fixed value for the irrelevant exponent:
$y_{i}=-0.83$ and the value $K_{c}=0.2216547$
($K=\frac{J}{K_{B}T}$) for the critical temperature. Thus, in
order to estimate the thermal critical exponent $y_{t}$, five more
parameters ($q_{o}, q_{1}, p_{o}, \textit{r}, s_{o}$) are involved
in (\ref{eq:17}). This ``many-parametric'' fit gave the estimate
$y_{t}=1.60(2)$, but the errors reported of all five parameters
were very large (up to 100\%) even for the coefficients
$q_{o}$($q_{1}$) of the leading singularity.

\begin{figure*}
\includegraphics*[width=12 cm]{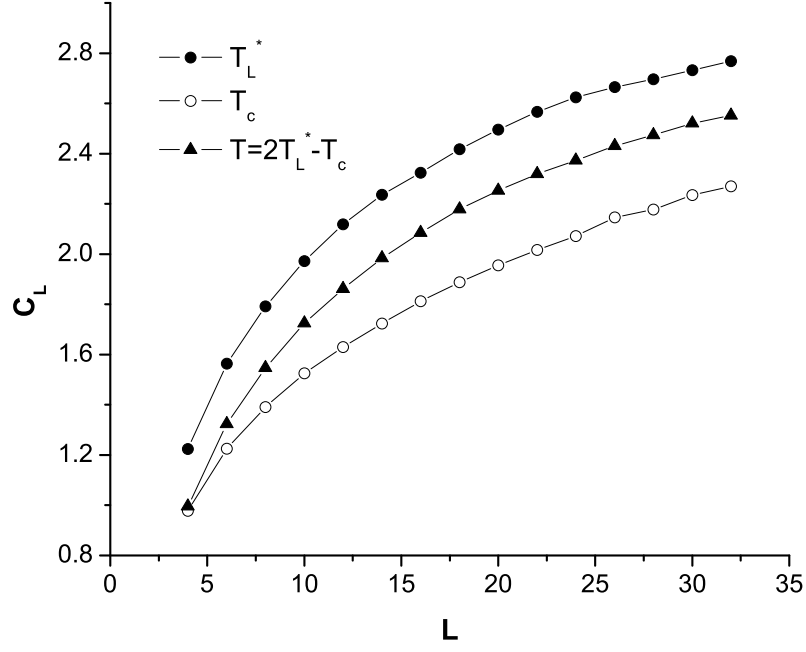}
\caption{\label{fig:fig3}Specific heat values (in units of
$k_{B}$) of the cubic Ising model for the three temperatures
mentioned in the text.}
\end{figure*}

We applied the CrMES Wang-Landau scheme to obtain the DOS for
lattice sizes $L=4-32$ for the simple cubic Ising lattice. For
each lattice, several random walks on the selected restricted
energy space were performed for averaging. The numbers of these
walks varied with the lattice size, ranging from $30$ walks for
the size $L=4$ to $100$ walks for the size $L=32$. We used the
same procedures for averaging and estimating the errors, described
in the previous section. In this way we obtained data for the
specific heat in the critical region following the method
described in Section II. We also attempted a similar analysis,
based on the expansion (\ref{eq:17}), fixing the irrelevant
exponent to the value $y_{i}=-0.821$ from Y. Deng and H. Blote
\cite{deng03}. In particular, we concentrated on three
temperatures: the ``pseudo-critical'' temperatures $T_{L}^{\ast}$,
a ``good'' approximation $T_{c}'=4.51152..$ ($K_{c}=0.2216547$
\cite{blote95}) for the exact critical temperature $T_{c}$, and
finally a ``lower'' temperature defined for each lattice by
$\widehat{T}_{L}=2T_{L}^{\ast}-T_{c}'$. Fig.\ref{fig:fig3} shows
the values of the specific heat at these temperatures as function
of the lattice size. In Table \ref{tab:table4} we present our
estimates for the pseudo-critical temperatures $T_{L}^{\ast}$, the
corresponding values of the specific heat $C_{L}^{\ast}$ and the
extensions of the critical minimum energy subspaces (CrMES)
$\Delta\widetilde{E}^{\ast}(\textit{r})$ for the three levels of
accuracy $\textit{r}=10^{-3}, 10^{-4}, 10^{-6}$. From
Fig.\ref{fig:fig3} one can observe a rather smooth behavior with
relatively small errors. The estimates for the lower temperature
$\widehat{T}_{L}$ seem to be the most accurate. However, our
attempt to fit these data in the expansion (\ref{eq:17}) produced
modest estimates for the thermal exponent and very large errors
for almost all other parameters. We found estimates of the same
order with those given in Blote et al \cite{blote95}, at least for
the dominant terms of the expansion, but such fittings are not
reliable since the errors in all coefficients are very large.

In order to suppress the errors we tried to omit further terms
from the expansion and we searched for stable forms as we
disregarded the smaller lattice sizes from the fittings. Thus, we
have observed the fittings, for several alternative truncations of
the expansion, in the following six successive intervals: $L=4-32,
6-32, ..., 14-32$. Among other possibilities, we kept (as
non-zero) only the correction terms with coefficients $q_{o}$ and
$\textit{r}$ in (\ref{eq:17}). The resulting estimates for the
thermal exponent shifts to lower values as we move to larger-size
intervals. Thus, although some of the estimates seem to be very
close to the expected value of the critical thermal exponent, the
overall behavior is rather unsettled producing estimates for
$y_{t}$ in a rather wide range: $1.56-1.62$. An explanation for
this behavior may be the following: as we move to larger lattice
sizes the relative contribution of the various correction terms is
changing and this make the analysis for these relatively small
sizes very sensitive. However, some quite acceptable exceptions
will be now mentioned: Consider, the specific heat values at the
temperature $T_{c}'=4.51152..$ and fix the value of the constant
contribution in the neighborhood of $p_{o}\cong-1.5$, then allow
$q_{o}$ and $\textit{r}$ to vary and make successive fittings for
all intervals from $L=4-32$ up to $L=14-32$. These six fittings
are very good and stable and produce estimates with very small
errors. They give approximately the same value for the thermal
exponent but also for the coefficients $q_{o}$ and $\textit{r}$.
This is true for even larger-size ranges but with larger errors.
Considering the mean and the standard deviation of these six
estimates (see Appendix) we find:
\begin{equation}
\label{eq:18}
 y_{t}=1.5878(31),\;\ q_{o}=2.08(6),\;\
\textit{r}=-0.43(20)
\end{equation}

The above values should be compared with the values given in Blote
et al \cite{blote95}. Our error limits are about $10$ times
smaller and our estimate for the thermal exponent is very close to
the value given by Y. Deng and H. Blote \cite{deng03}
($y_{t}=1.5868(3)$). The constant term and the main amplitude
$q_{o}$ are just marginally in agreement with the values in Blote
et al \cite{blote95} ($p_{o}=-0.8(7)$ and $q_{o}=1.5(5)$). This is
a good coincidence and we may speculate that this exceptional case
is very close to the exact result. Its appearance may be well
related to the absence of the term with coefficient $q_{1}$ in the
expansion, which for the other two temperatures may cause fitting
problems. Furthermore, a stable sequence of fittings using the
specific heat values at the lower temperature
$\widehat{T}_{L}=2T^{\ast}_{L}-T'_{c}$ is also given in Appendix.
This sequence produces estimates for the thermal exponent $y_{t}$,
comparable with that given in (\ref{eq:18}). Finally, note that we
may use the values for $T^{\ast}_{L}$ to estimate the critical
temperature $T_{c}$ and/or the critical exponent $\nu$ from
(\ref{eq:2b}). The fitting for the case $L=12-32$ provides good
values for both these critical parameters, without even using
correction terms. To obtain values comparable in accuracy, with
the best known estimates, a study of several different
thermodynamic quantities may be necessary (see for instance
\cite{ferre91}).

Let us now examine the verification of our proposal for the
scaling of the extensions of the CrMES. The estimates for these
extensions are included in Table IV. Once again one can observe
that the reported relative errors for these extensions (for the
three levels of accuracy) are significantly smaller (by a factor
of $10$) than those concerning the values of the specific heat. It
is also remarkable that for sizes up to $L=16$ there are no errors
at all for these extensions. Note that, even the restriction of
the energy space using the larger value of the accuracy level
($\textit{r}$) will not introduce errors in the specific heat,
larger than those generated from the Wang-Landau random walk.
Thus, if we minimize our requirements so that we only calculate
the value of the specific heat at the pseudo-critical temperature,
then the energy subspace needed is only $1/20$ of the total energy
space for the $32\times32\times32$ cubic lattice. The extended
energy ranges used for the estimation of the parameters in Table
\ref{tab:table4} and the values of the specific heat at the
temperatures $T_{c}'$, $T^{\ast}_{L}$ are summarized in the
Appendix (see Table \ref{tab:table6}).

When the extensions of the CrMES are subjected to a finite size
analysis using a many-parametric expansion as (\ref{eq:17}), we
again find modest estimates for the thermal exponents and large
errors in all other parameters. However, we have discovered that
the dominant contributions now correspond to the terms with
coefficients $q_{o}$ and $\textit{r}$ in (\ref{eq:17}).
Introducing a more convenient notation we assume that these
extension scale as:
\begin{equation}
\label{eq:19}
\Psi(r)\equiv\left(
\frac{\Delta\widetilde{E}^{\ast}_{r}}{L^{d/2}}\right)^{2}\simeq
q(r)L^{2y_{t}-d}+p(r)L^{2y_{t}-d+y_{i}}
\end{equation}
Table \ref{tab:table5} shows successive fits on the above form for
the three levels of accuracy. As previously the value of the
irrelevant exponent is fixed to the value $y_{i}=-0.821$, but no
other parameter is fixed. The last two fittings give close
agreement (almost to the third decimal place) with the best known
estimate of the thermal critical exponent \cite{deng03}. There is
a small shift of the estimated thermal critical exponent to a
lower value as we move to larger lattice sizes indicating possible
existence of further correction terms. This shift is similar, but
considerably smaller, with the one detected in our fittings for
the specific heat values at the pseudo-critical temperatures.
Thus, we can conclude that the proposed scaling law of the CrMES
introduced in this paper is correct and can be considered as a new
effective technique for estimating the thermal critical exponent.

\section{Conclusions}
We have presented a simple and efficient approximation scheme,
which greatly facilitates the application of Wang-Landau sampling
in large systems for the estimation of critical behavior. In
particular, we have applied our proposal to study the finite size
behavior of the specific heat for both square and cubic Ising
lattices. It has been shown that one needs only a relatively small
part of spectral degeneracies in order to obtain good estimation
of the specific-heat peaks. We have described the outline of an
algorithm for identifying this part of the total energy range.
Furthermore, a scaling law for the finite size behavior of the
extensions of the critical part of the minimum energy subspace
(CrMES) determined with the help of a predefined level of accuracy
was proposed. This scaling law has been verified for both models
studied in this paper and estimates of the thermal critical
exponent for the three-dimensional case were obtained through this
route. Also in the two-dimensional case the expected logarithmic
behavior was confirmed.

In this paper we have considered an important aspect of the
problem concerning the extraction of the critical behavior, by
employing finite-size scaling theory and the recent methods that
directly calculate the density of states of classical statistical
models. Future applications of the proposed scheme concern several
models, for which we may use the Wang-Landau technique or the
broad histogram \cite{ferre98,lima00} and transition matrix
\cite{wang02,wang99} methods. However, the main goal is to improve
accuracy and obtain high-quality data for substantially larger
lattices. This may be achieved now with the help of our proposal
but the need of a comprehensive examination of all ``systematic''
and statistical errors of the DOS methods is now indispensable.
The errors, for example, when implementing the Wang-Landau method
are coming from several sources. There are errors coming from the
finite accuracy of the histogram flatness which may propagate and
amplify through the process of connecting the energy ranges in a
multi-range approach. There are also errors stemming from the
incomplete detailed balance condition. As always, we may expect
errors from the random number generation and the usual statistical
fluctuations. An ``optimization'' of all these errors, seems to be
at this time quite demanding. The multi-range approach described
in Section II, that leaves unchanged the central subinterval
degeneracies, is only one ``ingredient'' of such an optimization.

\begin{acknowledgments}
This research was supported by the Special Account for Research
Grants of the University of Athens under Grant No. $70/4/4071$ and
EPEAEK/PYTHAGORAS $70/3/7357$.
\end{acknowledgments}

\section{Appendix}
Here we present specific heat values obtained by the proposed
CrMES Wang-Landau method and give further details of the fitting
attempts to the expansion (\ref{eq:17}) for the cubic Ising model.
Table \ref{tab:table6} gives the specific heat values and
specifies the extended energy subspace (CrMES) used in this paper
in order to obtain the accuracy level $r=10^{-6}$ and also
estimate the extensions given in Table \ref{tab:table4}. The
values $C(T_{c}')$ of the third column of Table \ref{tab:table6}
are now fitted in the following scaling formula:
\begin{equation}
\label{eq:20} C(T_{c}')=-1.5+q_{o}L^{2y_{t}-3}+rL^{2y_{t}-3.821}
\end{equation}
The successive estimates for the amplitudes $q_{o}$ and $r$ and
the thermal exponent $y_{t}$ are given in Table \ref{tab:table7}.
Their mean values over the fitting ranges appear in our proposal
in (\ref{eq:18}). Finally the values of $C(\widehat{T}_{L})$ are
fitted in a more restricted form (\ref{eq:21}), given below. The
produced estimates are shown in Table \ref{tab:table8}.
\begin{equation}
\label{eq:21}
C(\widehat{T}_{L})=-0.3+q_{o}L^{2y_{t}-3}-2L^{2y_{t}-3.821}
\end{equation}
The particular values of the expansion (\ref{eq:17}) for $p_{o}$
and $p_{o}$ and $r$ chosen in (\ref{eq:20}) and (\ref{eq:21})
respectively, provide a stable and convincing picture for the
estimation of the thermal exponent.

{}

\begin{table*}
\caption{\label{tab:table1}Exact and approximate Wang-Landau (WL)
results for the square Ising model. ``Pseudo-critical''
temperatures $T^{\ast}_L$, corresponding specific heat values
$C^{\ast}_L$ and specific heat values at the exact critical
temperatures $C_{L}(T_c)$. The exact DOS has been obtained by the
algorithm provided in \cite{beale96}.}
\begin{ruledtabular}
\begin{tabular}{lllllll}
$L$ &$T^{*}_L$: exact &$T^{*}_L$: WL &$C^{*}_L$: exact &$C^{*}_L$: WL &$C_L(T_c)$: exact &$C_L(T_c)$: WL\\
\hline 10 &2.34459 & &1.30906 & &1.26002 \\
 14 &2.32407 & &1.48356 & &1.43117 \\
 20 &2.30820 & &1.66628 & &1.61116 \\
 24 &2.30190 & &1.75899 & &1.70273 \\
 30 &2.29553 & &1.87193 & &1.81450 \\
 34 &2.29251 & &1.93507 & &1.87706 \\
 40 &2.28909 & &2.01686 & &1.95818 \\
 44 &2.28731 & &2.06473 & &2.00570 \\
 50 &2.28518 &2.2853(3) &2.12885 &2.1338(150) &2.06938
 &2.0791(150)\\
 54 & &2.2835(3) & &2.1690(200) & &2.1150(250)\\
 60 & &2.2825(3) & &2.2275(300) & &2.1620(350)\\
 64 & &2.2816(3) & &2.2523(350) & &2.1860(300)\\
 70 & &2.2809(3) & &2.2910(350) & &2.2242(350)\\
 74 & &2.2805(4) & &2.3240(400) & &2.2587(400)\\
 80 & &2.2796(4) & &2.3630(400) & &2.3000(450)\\
 84 & &2.2792(4) & &2.3900(450) & &2.3240(500)\\
 90 & &2.2781(4) & &2.4200(450) & &2.3650(500)\\
 94 & &2.2779(4) & &2.4460(500) & &2.3840(500)\\
 100 & &2.2775(4) & &2.4710(600) & &2.4120(500)\\
\end{tabular}
\end{ruledtabular}
\end{table*}

\begin{table*}
\caption{\label{tab:table2}Critical Minimum Energy Subspace
(CrMES) extensions for the square Ising model calculated for the
three predefined levels of accuracy $\textit{r}$. Note that the
relative errors for these extensions are much smaller than those
for the corresponding specific heats.}
\begin{ruledtabular}
\begin{tabular}{lllllllllllll}
$L$ &exact &WL &exact &WL &exact &WL &exact &WL &exact &WL &exact
&WL\\
&$\Delta\widetilde{E}^{*}(r_1)\footnotemark[1]$
&$\Delta\widetilde{E}^{*}(r_1)$
&$\Delta\widetilde{E}^{*}(r_2)\footnotemark[1]$
&$\Delta\widetilde{E}^{*}(r_2)$
&$\Delta\widetilde{E}^{*}(r_3)\footnotemark[1]$
&$\Delta\widetilde{E}^{*}(r_3)$
&$\Delta\widetilde{E}_c(r_1)$ &$\Delta\widetilde{E}_c(r_1)$ &$\Delta\widetilde{E}_c(r_2)$ &$\Delta\widetilde{E}_c(r_2)$ &$\Delta\widetilde{E}_c(r_3)$ &$\Delta\widetilde{E}_c(r_3)$\\
\hline 10 &38 & &40 & &45 & &36 & &39 & &43\\
 14 &63 & &68 & &74 & &62 & &66 & &72& \\
 20 &101 & &111 & &127 & &99 & &109 & &126 & \\
 24 &126 & &140 & &161 & &124 & &138 & &159 & \\
30 &165 & &184 & &213 & &163 & &182 & &211 & \\
34 &191 & &213 & &248 & &190 & &211 & &246 & \\
40 &232 & &259 & &302 & &230 & &257 & &300 & \\
44 &259 & &290 & &339 & &257 & &288 & &336 & \\
50 &301 &301 &337 &337 &394 &394 &299 &299 &335 &335 &392 &392 \\
54 & &329 & &368 & &432 & &327 & &366 & &429 \\
60 & &371 & &416 & &489 & &369 & &414 & &486 \\
64 & &400 & &448 & &527 & &398 & &446 & &524 \\
70 & &442(1) & &497 & &584(1) & &440(1) & &494(1) & &582 \\
74 & &472(1) & &530(1) & &624(1) & &470(1) & &528(1) & &621(1) \\
80 & &516(1) & &580(2) & &682(2)& &514(1) & &577(1) & &679(2) \\
84 & &545(1) & &613(2) & &722(2)& &543(1) & &610(2) & &719(2) \\
90 & &589(1) & &663(2) & &781(2)& &587(1) & &660(2) & &778(2) \\
94 & &619(2) & &696(2) & &821(2)& &616(2) & &693(2) & &818(2) \\
100 & &662(2) & &745(2) & &881(2)& &659(2) & &742(2) & &877(2) \\
\end{tabular}
\end{ruledtabular}
\footnotetext[1]{$r_1=10^{-3}$, $r_2=10^{-4}$ and $r_3=10^{-6}$}
\end{table*}

\begin{table*}
\caption{\label{tab:table3}Estimates of amplitudes obtained by
fitting (\ref{eq:16}) to the CrMES extensions presented in Table
\ref{tab:table2}\ for the square Ising model.}
\begin{ruledtabular}
\begin{tabular}{clll}
$L$ & &$A(r)$ &$B_1(r)$ \\
\hline &  $\Psi(r_1)$ &10.08(8) &-34(2) \\ $10-50$ &$\Psi_(r_2)$
&12.81(11) &-55(3) \\  &$\Psi(r_3)\footnotemark[1]$
&17.81(16) &-92(4)\\
\hline &  $\Psi(r_1)$ &9.90(3) &-32(1) \\ $10-100$ &$\Psi(r_2)$
&12.65(4)&-52(1) \\
& $\Psi(r_3)\footnotemark[1]$ &17.75(5)&-92(2)\\
\hline &  $\Psi(r_1)$ &9.83(3) &-29(2) \\ $50-100$ &$\Psi(r_2)$
&12.60(4)&-51(2)\\ &$\Psi(r_3)\footnotemark[1]$ &17.81(3)&-96(2)
\end{tabular}
\end{ruledtabular}
\footnotetext[1]{Mean value over the three fitting ranges:
$A(r_3)=17.80(6)$, $B_1(r_3)=-93(4)$}
\end{table*}

\begin{table*}
\caption{\label{tab:table4}Estimates obtained via Wang-Landau
CrMES scheme described in this paper  for the cubic Ising model.``
Pseudo-critical'' temperatures $T^{*}_L$, corresponding specific
heat values and CrMES extensions for the three levels of accuracy
$r$.}
\begin{ruledtabular}
\begin{tabular}{llllll}
$L$ &$T^{*}_L$ &$C(T^{*}_L)$ &$\Delta\widetilde{E}^{*}(r_1)\footnotemark[1]$ &$\Delta\widetilde{E}^{*}(r_2)\footnotemark[1]$ &$\Delta\widetilde{E}^{*}(r_3)\footnotemark[1]$\\
\hline 4 &4.1150(20) &1.2242(30) &45 &47 &51\\ 6
&4.2752(20)&1.5632(50) &122 &133 &148\\
8&4.3495(20)&1.7920(70)&215 &236 &269\\
10&4.3944(20)&1.9720(100)&325 &359 &413\\
12&4.4210(20)&2.1180(170)&453 &502 &579\\
14&4.4395(30)&2.2365(240)&596 &661 &766\\
16&4.4532(30)&2.3243(380)&753 &836 &971\\
18&4.4623(30)&2.4177(390)&925(1) &1029(2) &1195(2)\\
20&4.4702(30)&2.4955(400)&1106(1) &1232(2) &1433(2)\\
22&4.4758(30)&2.5670(420)&1302(2) &1451(2) &1690(2)\\
24&4.4796(30)&2.6242(450)&1504(6) &1677(6) &1954(10)\\
26&4.4854(35)&2.6648(750)&1722(8) &1926(8) &2244(10)\\
28&4.4879(45)&2.6966(800)&1951(8) &2178(8) &2544(10)\\
30&4.4911(45)&2.7325(900)&2178(10) &2432(10) &2845(10)\\
32&4.4929(50)&2.7688(900)&2430(10)&2716(10) &3176(12)\\
\end{tabular}
\end{ruledtabular}
\footnotetext[1]{$r_1=10^{-3}$, $r_2=10^{-4}$ and $r_3=10^{-6}$}
\end{table*}

\begin{table*}
\caption{\label{tab:table5}Fitting attempts using (\ref{eq:19}) to
estimate the thermal exponent $y_{t}$ from the CrMES extensions
shown in Table \ref{tab:table4} for the cubic Ising model. Note
that the mean values (given in the footnote below) for $y_{t}$ are
close to the value $y_{t}=1.5878(31)$ given in our proposal
(\ref{eq:18}) and to the value $y_{t}=1.5868(3)$ of
\cite{deng03}.}
\begin{ruledtabular}
\begin{tabular}{cllll}
$L$ & &$q(r)$ &$p(r)$ &$y_t$ \\
\hline &$\Psi(r_1)$ &108(2)&-261(7)&1.596(3)\\ $4-32$
&$\Psi_(r_2)$&132(3)&-328(10)&1.601(3)\\ &$\Psi(r_3)$
&170(3)&-440(11)&1.610(3) \\
\hline  &$\Psi(r_1)$ &107(4)&-255(14)&1.598(5)\\
$6-32$&$\Psi(r_2)$ &130(5)&-321(19)&1.602(5)\\
&$\Psi(r_3)$ &179(5)&-476(18)&1.603(3) \\
\hline  &$\Psi(r_1)$ &115(5)&-293(20)&1.588(5)\footnotemark[1]\\
$8-32$\footnotemark[1] &$\Psi(r_2)$
&142(6)&-373(27)&1.591(6)\footnotemark[1]\\
&$\Psi(r_3)$ &189(6)&-524(27)&1.596(4)\footnotemark[1] \\
\hline & $\Psi(r_1)$ &129(6)&-356(25)&1.574(6)\footnotemark[2]\\
$10-32$\footnotemark[2]&$\Psi(r_2)$
&155(9)&-439(40)&1.579(7)\footnotemark[2]\\ &$\Psi(r_3)$
&203(8)&-589(39)&1.587(6)\footnotemark[2]\\
\end{tabular}
\end{ruledtabular}
\footnotetext[1]{Mean value (over the three levels of accuracy) of
the thermal exponent $y_{t}=1.592(4)$} \footnotetext[2]{Mean value
(over the three levels of accuracy) of the thermal exponent
$y_{t}=1.580(7)$}
\end{table*}

\begin{table*}
\caption{\label{tab:table6}Specific heat values for the cubic
Ising model at the two temperatures $\widehat{T}_{L}$ and $T_{c}'$
defined in the text (see also footnote). The counting variables
$ie(E'_{min})$ and $ie(E'_{max})$, where $ie(E)=(E+3N)/4+1$,
specify the extended range used in this work. The portion of the
energy space used in our calculations is given in the last
column.}
\begin{ruledtabular}
\begin{tabular}{lllllc}
$L$ &$C(\widehat{T}_{L}\footnotemark[1])$ &$C(T_{c}')$
&$ie(E'_{min})$
&$ie(E'_{max})$ &$(E'_{max}-E'_{min})/(E_{max}-E_{min})$\\
\hline 4 &0.9954(20) &0.9776(20) &1 &70 &0.73\\
6&1.3230(30)&1.2256(30) &1 &170 &0.52\\
8&1.5471(60)&1.3908(60)&30 &380 &0.46\\
10&1.7245(80)&1.5248(80)&150 &670 &0.35\\
12&1.8610(80)&1.6300(80)&360 &1100 &0.29\\
14&1.9846(150)&1.7234(150)&710 &1670 &0.23\\
16&2.0854(250)&1.8129(250)&1220 &2410 &0.19\\
18&2.1790(350)&1.8883(350)&1800 &3500 &0.19\\
20&2.2527(380)&1.9555(380)&2800 &4520 &0.14\\
22&2.3190(380)&2.0170(380)&3900 &5960 &0.13\\
24&2.3728(400)&2.0720(400)&5250 &7650 &0.12\\
26&2.4307(470)&2.1458(470)&6870 &9680 &0.11\\
28&2.4738(500)&2.1772(600)&8790 &11970 &0.10\\
30&2.5209(650)&2.2345(800)&11150 &14590 &0.09\\
32&2.5515(700)&2.2700(850)&13700&17650 &0.08\\
\end{tabular}
\end{ruledtabular}
\footnotetext[1]{$\widehat{T}_{L}=2T^{\ast}_{L}-T_{c}'$,
$T_{c}'=4.51152..(K_{c}'=0.2216547)$}
\end{table*}

\begin{table*}
\caption{\label{tab:table7}Successive fittings for the specific
heat values $C(T_{c}')$. Scaling expansion (\ref{eq:20}) is used.}
\begin{ruledtabular}
\begin{tabular}{clll}
$L$ &$q_{o}$ &$r$ &$y_{t}\footnotemark[1]$ \\
\hline 4-32 &2.09(2) &-0.46(6) &1.5869(2)\\  6-32
&2.04(4)&-0.27(10) &1.5902(22)\\
8-32&2.03(6)&-0.27(17)&1.5904(33)\\
10-32&2.03(9)&-0.27(28)&1.5904(50)\\
12-32&2.11(13)&-0.54(45)&1.5860(71)\\
14-32&2.17(2)&-0.76(74)&1.5828(105)\\
\end{tabular}
\end{ruledtabular}
\footnotetext[1]{mean value $y_{t}=1.5878(31)$}
\end{table*}

\begin{table}
\caption{\label{tab:table8}Successive fittings for the specific
heat values $C(\widehat{T}_{L})$. The expansion used is given in
(\ref{eq:21}).}
\begin{ruledtabular}
\begin{tabular}{cll}
$L$ &$q_{o}$ &$y_{t}\footnotemark[1]$ \\
\hline 4-32 &1.633(7) &1.592(1)\\
6-32 &1.625(9)&1.593(1)\\ 8-32&1.627(11)&1.593(1)\\
10-32&1.637(13)&1.592(1)\\ 12-32&1.650(15)&1.590(1)\\
14-32&1.671(15)&1.588(2)\\ 16-32&1.692(16)&1.586(2)\\
18-32&1.718(14)&1.584(1)\\ 20-32&1.733(17)&1.582(2)\\
\end{tabular}
\end{ruledtabular}
\footnotetext[1]{mean value $y_{t}=1.5889(41)$}
\end{table}


\begin{thebibliography}{}
\bibitem{metro53} N. Metropolis, A. W. Rosenbluth, M. N. Rosenbluth, A. H. Teller
and E. Teller, J. Chem. Phys. {\bf 21}, 1087, (1953).

\bibitem{glaub63} R. J. Glauber, J. Math. Phys. {\bf 4}, 294, (1963).

\bibitem{bortz75} A. B. Bortz, M. H. Kalos and J. L. Lebowitz, J. Comput. Phys. {\bf 17},
10, (1975).

\bibitem{binde77} K. Binder, Rep.Prog. Phys. {\bf 60}, 487,
(1977).

\bibitem{newma99} M. E. J. Newman and G. T. Barkema, \textit{Monte Carlo Methods in
Statistical Physics}, Clarendon Press, Oxford, (1999).

\bibitem{landa00} D. P. Landau and K. Binder, \textit{A Guide to Monte Carlo Simulations in
Statistical Physics}, Cambridge University Press, (2000).

\bibitem{lee93} J.Lee, Phys. Rev. Lett. {\bf 71}, 211, (1993).

\bibitem{berg92} B. A. Berg and T. Neuhaus, Phys. Rev. Lett. {\bf 68}, 9, (1992).

\bibitem{ferre98} A. M. Ferrenberg and R. H. Swendsen, Phys. Rev. Lett. {\bf 61},
2635, (1998).

\bibitem{lima00} A. R. Lima, P. M. C. de Oliveira, and T. J. P.
Penna, J. Stat. Phys. {\bf 99}, Nos.3/4, (2000).

\bibitem{wang02} J. -S. Wang and R. H. Swendsen, J. Stat. Phys. {\bf
106}, Nos. 1/2, (2002).

\bibitem{wang99} J. -S. Wang, T. K. Tay and R. H. Swendsen, Phys. Rev. Lett. {\bf
82}, 476, (1999).

\bibitem{wang01} F. Wang and D. P. Landau, Phys. Rev. Let. {\bf 86}, 2050, (2001).

\bibitem{wang01b} F. Wang, D. P. Landau, Phys. Rev. E {\bf 64}, 056101, (2001).

\bibitem{schul01} B. J. Schulz, K. Binder and M. Muller, Int. J. Mod.
Phys. C {\bf 13}, 477, (2001).

\bibitem{schul03} B. J. Schulz, K. Binder, M. Muller and D. P. Landau, Phys. Rev.
E {\bf 67}, 067102, (2003).

\bibitem{landa02} D. P. Landau and F. Wang, Comput. Phys. Comm. {\bf 147}, 674, (2002).

\bibitem{ferdi69} A. E. Ferdinand and M. E. Fisher, Phys. Rev. {\bf 185}, 832, (1969).

\bibitem{fishe71} M. E. Fisher, \textit{Critical Phenomena}, ed. M. S. Green,
Academic Press, London, (1971).

\bibitem{privm90} V. Privman, \textit{Finite Size Scaling and Numerical Simulation of
Statistical Systems}, World Scientific, Singapore, (1990).

\bibitem{binde92} K. Binder, \textit{Computational Methods in Field Theory},
eds. C. B. Lang and H   Gausterer, Springer, Berlin, (1992).

\bibitem{beale96} P. de Beale, Phys. Rev. Lett. {\bf 76}, 1, (1996).

\bibitem{ferre91} A. M. Ferrenberg and D. P.Landau, Phys.Rev. B {\bf 44}, 10,
(1991).

\bibitem{fishe67} M. E. Fisher, Rep. Prog. Phys. {\bf 30}, 615, (1967).

\bibitem{domb60} C. Domb, Adv. Phys. {\bf 9}, 149, (1960).

\bibitem{nicke80} B. G. Nickel, \textit{Phase Transitions}: Cargese 1980 (Plenum, New
York, 1982), 291.

\bibitem{brezi74} E. Brezin, J. -C. Le Guillou and J. Zinn-Justin, Phys. Lett.
{\bf 47A}, 285, (1974).

\bibitem{guill80} J. -C. Le Guillou and J. Zinn-Justin, Phys. Rev. B {\bf 21},
3976, (1980).

\bibitem{guill87} J. -C. Le Guillou and J. Zinn-Justin, J. Phys. (Paris) {\bf 48},
19, (1987).

\bibitem{pawle84} G. S. Pawley, R. H. Swendsen, D. J. Wallace and K. G. Wilson
Phys. Rev. B {\bf 29}, 4030, (1984).

\bibitem{blote89} H. W. J. Blote, J. de Bruin, A. Compagner, J. H. Crookewit, Y.
T. J. C. Fonk, J. R. Heringa, A. Hoogland, and A. L. van Willigen
Europhys. Lett. {\bf 10}, 105, (1989).

\bibitem{blote89b} H. W. J. Blote, A. Compagner, J. H. Crookewit, Y. T. J. C.
Fonk, J. R. Heringa, A. Hoogland, T. S. Smit, and A. L. van
Willigen, Physica A {\bf 161}, 1, (1989).

\bibitem{liu89} A. J. Liu and M. E. Fisher, Physica A {\bf 165}, 35, (1989).

\bibitem{landa76} D. P. Landau, Phys. Rev. B {\bf 14}, 255, (1976).

\bibitem{barbe85} M. N. Barber, R. B. Pearson, D. Toussaint, and J. L.
Richardson, Phys. Rev. B {\bf 32}, 1970, (1985).

\bibitem{paris85} G. Parisi and F. Rapuano, Phys. Lett. {\bf 157B}, 301, (1985).

\bibitem{hoogl85} A. Hoogland, A. Compagner, and H. W. J. Blote, Physica A {\bf 132},
593, (1985).

\bibitem{bhano86} G. Bhanot, D. Duke, and R. Salvador, Phys. Rev. B {\bf 33},
7841, (1986).

\bibitem{chen82} J. H. Chen, M. E. Fisher and B. G. Nickel, Phys. Rev. Lett.
{\bf 48}, 630, (1982).

\bibitem{guttm94} A. J. Guttmann and I. G. Enting, J. Phys. A {\bf 27}, 8007, (1994).

\bibitem{koles94} M. Kolesik and M. Suzuki, Pysica A {\bf 27}, 8007, (1994).

\bibitem{nicke90} B. G. Nickel and J. J. Rehr, J. Stat. Phys. {\bf 61}, 1,
(1990).

\bibitem{nicke91} B. G. Nickel, Physica A {\bf 177}, 189,
(1991).

\bibitem{adler83} J. Adler, J. Phys. A {\bf 16}, 3585,
(1983).

\bibitem{talap96} A. L. Talapov and H. W. J. Blote, J. Phys. A {\bf 29}, 5727, (1996).

\bibitem{blote95} H. W. J. Blote, E. Luijten, and J. R. Heringa, J. Phys. A {\bf 28},
6289, (1995).

\bibitem{blote96b} H. W. J. Blote, J. R. Heringa, A. Hoogland, E. W. Meyer and
T. S. Smit, Phys. Rev. Lett. {\bf 76}, 2613, (1996).

\bibitem{deng03} Y. Deng and H. W. J. Blote, Phys. Rev. E {\bf 68},
036125, (2003).

\bibitem{blote97} H. W. J. Blote, L. N. Shchur, and A. L. Talapov, Int. J. Mod.
Phys. C {\bf 10}, 1137, (1997).

\bibitem{hanse91} M. Hansebusch, K. Pinn, and S. Vinti, Phys. Rev. B {\bf 59},
11471, (1991).

\bibitem{buter97} P. Butera and M. Comi, Phys. Rev. B {\bf 56}, 8212, (1997).

\bibitem{guida98} R. Guida and J. Zinn-Justin, J. Phys. A {\bf 31}, 8103, (1998).

\bibitem{garci} J. Garcia, J. A. Gonzalo and M. I. Marques, e-print
cond-mat/0211270.

\bibitem{buter02} P. Butera and M. Comi, Phys. Rev. B {\bf 65}, 144431,
(2002).

\end{thebibliography}
\end{document}